\documentclass[twocolumn,floatfix,superscriptaddress,showpacs]{revtex4}

\usepackage{graphicx}
\usepackage{amsmath}
\usepackage{epsfig}

\newcommand{\pa}{\partial}

\newcommand{\be}{\begin{equation}}
\newcommand{\e}{\end{equation}}
\newcommand{\beml}{\begin{subequations}}
\newcommand{\eml}{\end{subequations}}
\newcommand{\beq}{\begin{eqnarray}}
\newcommand{\eq}{\end{eqnarray}}
\newcommand{\ba}{\begin{array}}
\newcommand{\ea}{\end{array}}
\newcommand{\lt}{\left}
\newcommand{\rt}{\right}

\newcommand{\ep}{\varepsilon}

\begin{document}

\title{Photon-assisted electron transport in graphene}
\author{B. Trauzettel}
\affiliation{Instituut-Lorentz, Universiteit Leiden, P.O. Box 9506,
2300 RA Leiden, The Netherlands}
\affiliation{Department of Physics and Astronomy, University of Basel, 
Klingelbergstrasse 82, 4056 Basel, Switzerland}
\author{Ya. M. Blanter}
\affiliation{Kavli Institute of Nanoscience, Delft University of
Technology, Lorentzweg 1, 2628 CJ Delft, The Netherlands}
\author{A. F. Morpurgo}
\affiliation{Kavli Institute of Nanoscience, Delft University of
Technology, Lorentzweg 1, 2628 CJ Delft, The Netherlands}
\date{September 2006}
\begin{abstract}
Photon-assisted electron transport in ballistic graphene is
analyzed using scattering theory. We show that the presence of an
ac signal (applied to a gate electrode in a region of the system)
has interesting consequences on electron transport in graphene,
where the low energy dynamics is described by the Dirac equation.
In particular, such a setup describes a feasible way to probe energy 
dependent transmission in
graphene. This is of substantial interest because the energy
dependence of transmission in mesoscopic graphene is the basis of 
many peculiar transport phenomena proposed in the recent literature. 
Furthermore, we
discuss the relevance of our analysis of ac transport in graphene
to the observability of {\it zitterbewegung} of electrons that behave as
relativistic particles (but with a lower effective speed of light).
\end{abstract}
\maketitle

\section{Introduction}

\begin{figure}
\vspace{0.5cm}
\begin{center}
\epsfig{file=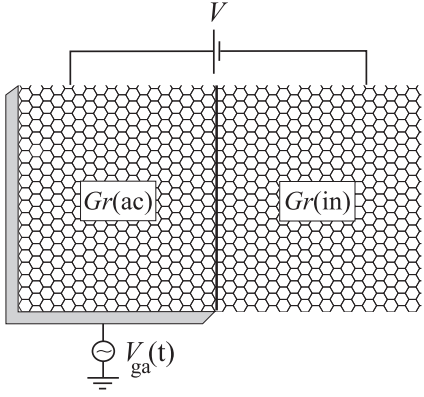,scale=0.45} \caption{\label{tg_setup}
Schematic of the setup. Two regions of graphene (called $Gr(\rm
ac)$ and $Gr(\rm in)$) are voltage-biased with respect to each
other with a bias $V$. A gate electrode (grey-shaded area) shifts
the Dirac point of region $Gr(\rm ac)$ away from zero energy.
Additionally a small ac signal is applied to the gate to generated
side band transitions. The applied gate
voltage $V_{\rm ga}(t)$ has to be chosen in such a way that the
(time-dependent) chemical potential in region $Gr(\rm ac)$ is
given by Eq.~(\ref{mu}).}
\end{center}
\end{figure}

Since the discovery of an anomalous quantum Hall effect in
graphene \cite{Nov05,Zha05}, promising possibilities to observe
quantum dynamics of Dirac fermions in such systems have been
proposed. The most prominent one is the quantum-limited
conductivity (of order $e^2/h$), where measurements
\cite{Nov05,Zha05} and theoretical predictions
\cite{Lud94,Zie98,Per05,Kat05,Nom06,Ost06} are still inconsistent with each
other. Other examples of unusual quantum transport phenomena in
mesoscopic graphene are a maximum Fano factor of $\mathbf{\frac{1}{3}}$
\cite{Two06}, selective transmission of Dirac electrons through
\textit{n-p} junctions \cite{Che06}, the phenomenon of Klein
tunnelling \cite{Kat06}, and transport phenomena at interfaces
between graphene and a superconductor \cite{Bee06,Tit06a,Cue06,Tit06b}.
Recently, the effect on the longitudinal and Hall conductivity of
an applied microwave signal has been analyzed in bulk graphene
\cite{Gus06}. In many of these works, interesting phenomena arise because
transport through graphene is in general energy dependent. Therefore,
it is desirable to directly probe the energy dependence of transport.
We show that the Tien-Gordon problem \cite{Tie63,Pla04}
of photon-assisted electron
transport for Dirac electrons is a powerful tool to
quantitatively probe energy dependent transmission in graphene. The idea
is to apply an ac signal to a region of graphene and a bias with respect
to a neighboring region to allow electron transport from the
one to the other (see Fig.~\ref{tg_setup}) \cite{Foot1}. 

In the dc limit, electron transport then allows to directly determine
the energy dependent transmission coefficients of the underlying scattering
problem. Additionally, we find that resonance phenomena 
(sharp steps in $dG/dV$, where
$G=dI/dV$ is the differential conductance) arise if
the applied bias $V$ equals multiples of the applied ac frequency
$\omega$. These resonance phenomena are due to the vanishing of
propagating modes directly at the Dirac point (the point in the
spectrum of graphene, where the valence band and the conduction
band touch each other). This implies an interesting application of
graphene: It can be used as a spectrometer for high frequency
noise -- similar to the superconductor--insulator--superconductor
(SIS) junction in Ref.~\cite{Deb03} with the advantage that there
is no frequency limit.

Another motivation to look at the Tien-Gordon problem in graphene
is its relevance to the observability of an interesting and unobserved
phenomenon, present for free Dirac fermions but absent for free
Schr{\"o}dinger fermions, called {\em zitterbewegung} (ZB)
\cite{Sch30,Hua52}. ZB manifests itself, for instance, in a
time-dependent oscillation of the position operator of a Dirac
electron in the Heisenberg picture. A pioneering attempt to
describe an experimental way to observe ZB in III--V semiconductor
quantum wells has been proposed in Ref.~\cite{Sch05}. Others have
adopted this idea to carbon nanotubes \cite{Zaw05}, spintronic,
graphene, and superconducting systems \cite{Cse06}. Another situation where
non-relativistic electrons experience ZB is the presence of an
external periodic potential \cite{Rus06}. In fact, there is a wide class
of Hamiltonians (where the corresponding wave function is a spinor)
which all exhibit ZB \cite{Win06}. Nevertheless -- although almost
omnipresent -- the phenomenon has never been observed in nature.
Therefore, it is desirable to propose an unambiguous signature of ZB
in a particular setup. As we show below, the conditions to
observe ZB can be easily created using the Tien-Gordon effect, which results
in an oscillating current due to the existence of a wavefunction that is 
a superposition of different energies. However, an unambiguous
detection of ZB in this way is complicated, since also for the free
Schr{\"o}dinger equation an oscillating current is  
generated when the wavefunction is a superposition of different
energies. Additional studies are needed to single out effects of ZB
and ensure their unambiguous observation.

The paper is organized as follows: In Sec.~\ref{sec_setup}, we introduce the
setup under consideration and the model to describe it. The photon-assisted
current is analyzed in Sec.~\ref{sec_pat}. Afterwards, in Sec.~\ref{sec_zb}, 
we discuss the relevance of the previous analysis to the observability of ZB
in graphene. Finally, we conclude in Sec.~\ref{sec_con}.

\section{Setup and model}
\label{sec_setup}

The setup under consideration is illustrated in
Fig.~\ref{tg_setup}. It contains two regions of graphene called
$Gr(\rm ac)$ and $Gr(\rm in)$. A gate electrode in region $Gr({\rm
ac})$ shifts the Dirac points of the two regions with respect to
each other by an amount $eV_S$. Additionally, we apply a small ac
signal to the gate electrode.

An ideal sheet of graphene (in the absence
of $K$--$K'$ inter-valley mixing) can be described in the effective
mass approximation by a two-dimensional Dirac equation (DE) for a two-component
wave-function envelope $\Psi=(\Psi_1,\Psi_2)$ (subscript $1,2$
refers to pseudospins, whose origin can be traced to the presence
of two carbon sublattices)
\be \label{Dirac_eq_e}
\left[ -i v \hbar \lt(\ba{cc} 0 & \pa_x -i \pa_y \\
\pa_x +i \pa_y & 0 \ea\rt) - \mu(x) \right] \Psi_{e/h} = i \hbar \partial_t
\Psi_{e/h} ,
\e
where $v$ is the Fermi velocity. The dimensions of the
sample are assumed to be large enough such that boundary effects
can be neglected. It is, however, straightforward to include boundary effects
along the lines of Ref.~\cite{Two06}. The index $e/h$ refers to electron-like
(energy $\ep>0$ with respect to the Dirac point) and hole-like
(energy $\ep<0$ with respect to the Dirac point) solutions to the
DE. The chemical potential of the two regions $Gr(\rm ac)$ ($x<0$)
and $Gr(\rm in)$ ($x>0$) is
\be \label{mu}
\mu(x) = \lt\{ \ba{cc} eV_S + eV_{\rm ac} \cos (\omega t) \; , & {\rm for}
\; x<0 \\ 0  \; , & {\rm for}
\; x>0 \ea\rt. \; .
\e
The potential $\mu(x)$ is chosen such that a solution to the DE with
components of different energies is generated in $Gr(\rm ac)$ and transmitted
to $Gr(\rm in)$. The
solution to the wave equation (\ref{Dirac_eq_e}) for $x<0$ (in
region $Gr(\rm ac)$) may then be written as
\begin{eqnarray} \label{state_ac}
 \Psi^{({\rm ac})}_{e/h} (\vec{x},t) &=& \Psi^{({\rm ac})}_{0,e/h} (\vec{x},t)
e^{-i(e V_{\rm ac}/\hbar \omega) \sin (\omega t)} \\
&=& \sum_{m=-\infty}^\infty J_m \left( \frac{e V_{\rm ac}}{\hbar \omega}
\right) \Psi^{(ac)}_{0,e/h} (\vec{x},t) e^{-i m \omega t} \nonumber ,
\end{eqnarray}
where $\Psi^{({\rm ac})}_{0,e/h} (\vec{x},t) =  \Psi^{({\rm ac})}_{0,e/h} (\vec{x})
e^{\mp i \ep t/\hbar}$ and
\be \label{Dirac_eq_e_Ib}\left[
-i v \hbar \lt(\ba{cc} 0 & \pa_x -i \pa_y \\
\pa_x +i \pa_y & 0 \ea\rt) - eV_S \right]
\Psi^{({\rm ac})}_{0,e/h} = \pm \ep
\Psi^{({\rm ac})}_{0,e/h} .
\e
(We have dropped the argument $\vec{x}$ of the wave function in the
latter equation.) In
Eq.~(\ref{state_ac}), $J_m$ is the $m$th order Bessel function.
>From now on, we focus without loss of generality on
electron-like solutions only and drop the index $e$. This is justified
because, in ballistic transport from one region of graphene to another
region, particles with a fixed energy $\ep$ (which can be either positive or
negative with respect to the Dirac point) are transmitted and their
energy is conserved in the absence of inelastic scattering.

The electron-like plane wave solutions of Eq.~(\ref{Dirac_eq_e_Ib}) can be
written as linear combinations of the basis states \cite{Bee06}
\beq \Psi^{({\rm ac})}_{0,+} &=& \frac{e^{i q y + ik_{\rm ac} x}}
{\sqrt{\cos \alpha_{\rm ac}}}
\lt( \ba{c} e^{-i \alpha_{\rm ac}/2} \\ e^{i \alpha_{\rm ac}/2} \ea \rt) ,
\label{b1e} \\
\Psi^{({\rm ac})}_{0,-} &=& \frac{e^{i q y - ik_{\rm ac} x}}{\sqrt{\cos \alpha_{\rm ac}}}
\lt( \ba{c} e^{i \alpha_{\rm ac}/2} \\ - e^{- i \alpha_{\rm ac}/2} \ea \rt) ,
\label{b2e}
\eq
where
\begin{equation} \label{alphaac}
\alpha_{\rm ac} = \arcsin \left( \frac{\hbar v q}{\ep + eV_S} \right) ,
\end{equation}
$q$ is the transversal momentum, and $k_{\rm ac}=(\ep + eV_S)\cos
\alpha_{\rm ac}/(\hbar v)$ the longitudinal momentum in region
$Gr(\rm ac)$. Likewise, electron-like plane wave solutions of
Eq.~(\ref{Dirac_eq_e}) in region $Gr(\rm in)$ can be written as
linear combinations of the basis states
\beq \Psi^{({\rm in})}_{0,+} &=& \frac{e^{i q y + ik_{\rm in} x}}
{\sqrt{\cos \alpha_{\rm in}}}
\lt( \ba{c} e^{-i \alpha_{\rm in}/2} \\ e^{i \alpha_{\rm in}/2} \ea \rt) ,
\label{b1ebis} \\
\Psi^{({\rm in})}_{0,-} &=& \frac{e^{i q y - ik_{\rm in} x}}{\sqrt{\cos \alpha_{\rm in}}}
\lt( \ba{c} e^{i \alpha_{\rm in}/2} \\ - e^{- i \alpha_{\rm in}/2} \ea \rt) ,
\label{b2ebis}
\eq
with
$\alpha_{\rm in} = \arcsin (\hbar v q/\ep)$ and $k_{\rm
in}=(\ep/\hbar v) \cos \alpha_{\rm in}$. Now, we solve the
transmission problem from region $Gr(\rm ac)$ to region $Gr(\rm
in)$. An incoming wave function from region $Gr(\rm ac)$ is given
by
\be \label{psiin}
\Psi^{(\rm ac)}_{i} (\vec{x},t) = \sum_{m=-\infty}^\infty
J_m \left( \frac{e V_{\rm ac}}{\hbar \omega}
\right) \Psi^{(\rm ac)}_{0,+} e^{-i(\ep + \hbar m \omega)
t/\hbar} . \nonumber
\e
The reflected wave function in region $Gr(\rm ac)$ reads
\be \label{psire}
\Psi^{(\rm ac)}_{r} (\vec{x},t) = \sum_{m=-\infty}^\infty r_m
J_m \left( \frac{e V_{\rm ac}}{\hbar \omega}
\right) \Psi^{(\rm ac)}_{0,-} e^{-i (\ep + \hbar m \omega)
t/\hbar} , \nonumber
\e
where $r_m$ is the energy-dependent reflection coefficient.
Furthermore, a transmitted wave function in region
$Gr(\rm in)$ can be written as
\be \label{psitr}
\Psi^{(\rm in)}_{\rm tr} (\vec{x},t) = \sum_{m=-\infty}^\infty
t_{m} \ J_m \left( \frac{e V_{\rm ac}}{\hbar \omega}
\right) \Psi^{(\rm in)}_{+,m} e^{-i (\ep + \hbar m \omega)
t/\hbar} ,
\e
where
\be
 \Psi^{(\rm in)}_{+,m} (\vec{x}) =
\frac{e^{i q y + ik_{{\rm in},m} x}}{\sqrt{\cos \alpha_{{\rm in},m}}}
\lt( \ba{c} e^{-i \alpha_{{\rm in},m}/2} \\ e^{i \alpha_{{\rm in},m}/2}
\ea \rt)
\e
with
\begin{equation} \label{alphain}
\alpha_{{\rm in},m} = \arcsin \left( \frac{\hbar v q}{\ep +
\hbar m \omega} \right)
\end{equation}
and
$k_{{\rm in},m} = (\ep + \hbar m \omega)\cos\alpha_{{\rm in},m}/(\hbar v)$.
In Eq.~(\ref{psitr}), $t_{m}$ is the energy-dependent transmission coefficient.
In order to determine $t_m$ and $r_m$, we need to match wave functions at
$x=0$, namely
\[
\Psi^{(\rm ac)}_{i}(x=0,y,t) + \Psi^{(\rm ac)}_{r}(x=0,y,t) =
\Psi^{(\rm in)}_{\rm tr}(x=0,y,t) \nonumber .
\]
The solutions to the resulting set of equations are
\beq \label{randt}
r_{m} &=& \frac{e^{i \alpha_{\rm ac}} - e^{i \alpha_{{\rm in},m}}}
{1+e^{i(\alpha_{\rm ac} + \alpha_{{\rm in},m})}} , \\
t_{m} &=& e^{-i(\alpha_{\rm ac}-\alpha_{{\rm in},m})/2}
\sqrt{\frac{\cos \alpha_{{\rm in},m}}{\cos \alpha_{\rm ac}}}
\frac{1+e^{2i\alpha_{\rm ac}}}{1+e^{i(\alpha_{\rm ac}
+\alpha_{{\rm in},m})}} , \nonumber
\eq
where $\alpha_{\rm ac}$ is given by Eq.~(\ref{alphaac}) and
$\alpha_{{\rm in},m}$ by Eq.~(\ref{alphain}).
It is easy to verify that unitarity holds $|r_{m}|^2 + |t_{m}|^2 = 1$.
Furthermore, Eq.~(\ref{randt}) shows
that if the angle of incidence is zero ($q=0$) then all transmission
coefficients are $1$ and all reflection coefficients vanish. This is
known as Klein tunnelling in relativistic quantum dynamics \cite{Kat06}.

\section{Photon-assisted current}
\label{sec_pat}

In order to determine the transmitted current, we need to calculate the
current density operator in $x$-direction
\begin{equation}
J_x(\vec{x},t) = e v  \Psi^*(\vec{x},t) \sigma_x \Psi(\vec{x},t) ,
\end{equation}
integrate over a cross section in $y$-direction, and over angles of incidence.
In the following, we assume that a dc bias $V$ is applied between regions
$Gr(\rm ac)$ and $Gr(\rm in)$ and that $k_B T$ is the lowest of all energy
scales. The average current is then given by
\beq \label{curr1}
I &=& \frac{4e}{h} \frac{W}{\pi} \int_0^{q_{\rm max}} dq \int_0^{eV} d \ep
\sum_{m,m'=-\infty}^\infty \\
&\times& J_m \left( \frac{e V_{\rm ac}}{\hbar \omega} \right)
J_{m'} \left( \frac{e V_{\rm ac}}{\hbar \omega} \right) t^*_{m} t_{m'}
e^{i(m-m')\omega t} , \nonumber
\eq
where $W$ is the width of the sample and $q_{\rm max} = {\rm min}
\Bigl\{ |eV + \hbar m \omega|/\hbar v, |eV+ eV_S|/\hbar v \Bigr\}$
is the upper bound of the transversal momentum of propagating
modes. The bias is applied such that the Fermi function
(at zero temperature) in region $Gr(\rm ac)$ reads $f_{\rm ac}(E)
= \theta(eV-E)$ and, in region $Gr(\rm in)$, $f_{\rm in}(E) =
\theta(-E)$, where $\theta(x)$ is the Heaviside step function.
Since the scattering matrix is energy-dependent, the current $I$
depends on the way the bias is applied and not just on its
magnitude $V$. A factor $4$ has been added to the right hand
side of Eq.~(\ref{curr1}) to take into account for spin and valley
degeneracy. Apparently, there are terms in Eq.~(\ref{curr1}) that
do not depend on time (where $m=m'$) and terms that oscillate as a
function of time (where $m \neq m'$).

We will now show that the dc limit of Eq.~(\ref{curr1}) 
contains all the information of the energy dependent transmission 
of the underlying relativistic quantum dynamics. 
In that limit, only terms with $m=m'$ survive in
Eq.~(\ref{curr1}) and the differential conductance reads
\be \label{dIdV}
G = \frac{4 e^2}{h} \frac{W}{\pi} \int_0^{q_{\rm max}} dq
\sum_{m= -\infty}^\infty J^2_m
\left( \frac{e V_{\rm ac}}{\hbar \omega} \right) |t_{m} (\ep = eV)|^2 .
\e
%
\begin{figure}
\vspace{0.5cm}
\begin{center}
\epsfig{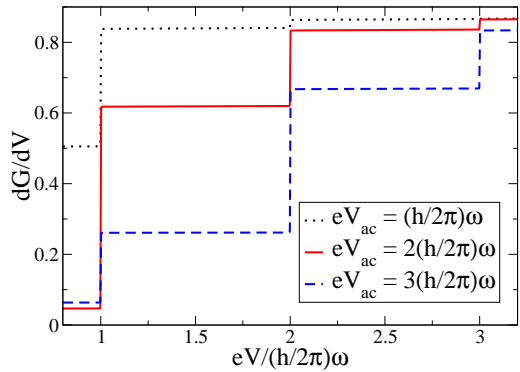}
\caption{\label{fig2} (Color online) $dG/dV$ is plotted
in units of $8 e^3 W/(vh^2)$ for a fixed value of
$eV_S/\hbar \omega = 100$ and different values of $eV_{\rm ac}/\hbar \omega$.
Sharp steps appear when the applied bias equals multiples of the ac frequency.
The size of the steps is non-universal. It depends on the weight of the
different side bands given by the Bessel functions in Eq.~(\ref{dIdV}). For
the given values of $eV_{\rm ac}/\hbar \omega$, it is sufficient to include
side bands up to $|m|=4$ to calculate $dG/dV$ to a very high accuracy.}
\end{center}
\end{figure}
The latter equation shows that the combination of the presence of the bias
$V$ and the ac signal $V_{\rm ac}$ allows to extract the energy-dependence
of the transmission coefficients (in principle) channel by channel. 
Typical values for $\omega$ that can be used experimentally in practice are
up to $10 - 30$ GHz. The bias can be controlled with any desired precision
relative to $k_B T$ down to mK temperature.

Furthermore, we find that $G$ has an interesting feature which might 
give rise to
potential applications. As illustrated in
Fig.~\ref{fig2}, steps appear in $dG/dV$ if the bias $V$ equals
multiples of the ac frequency $\omega$. The steps show up whenever
the number of propagating modes in a side band $m$ vanishes. That
is when the corresponding energy of charge carriers hits the Dirac
point, where the density of states vanishes. The magnitude of the
steps depends on the weight of the different side bands (given by
the Bessel functions) and is non-universal. Nevertheless, the
appearance of these steps can be used as a sensitive detector for
finite frequency noise, similar to the SIS junction in
Ref.~\cite{Deb03}. The advantage of graphene as a finite frequency
noise detector is that it has no frequency limit,
whereas the SIS detector is limited by the size of the
superconducting gap.

At this point, we mention that our results are obtained using scattering
theory of non-interacting Dirac fermions. As shown by Pedersen and B{\"u}ttiker
\cite{Ped98} (for the corresponding problem based on
the Schr{\"o}dinger equation)
screening of near-by gates can change the predictions of a scattering theory
based on non-interacting particles. Although screening is expected to
be reduced close to the Dirac point in graphene \cite{Div84}, where the
interesting predictions illustrated in Fig.~\ref{fig2} appear,
a careful analysis of mesoscopic ac transport in graphene in the presence of
screening by near-by gates is an interesting research project by itself.

\section{Relation to zitterbewegung}
\label{sec_zb}

Let us first explain the signature of ZB in the current in the
absence of an oscillating potential (as introduced in Eq.~(\ref{mu})) and 
then show why our particular choice of the oscillating potential has
some relevance as far as ZB is concerned. In the absence of the potential
$\mu(x)$, the time evolution of the electron field operator that
obeys Eq.~(\ref{Dirac_eq_e}) can be written as
\be \label{des_state}
\Psi_{\bf p}(t) = \frac{1}{2} \left[ \Psi^{(+)}_{\bf p}(t) +
\Psi^{(-)}_{\bf p}(t) \right]
\e
with ${\bf p} = (p_x,p_y)$, $p=\sqrt{p_x^2+p_y^2}$, and
\[
\Psi^{(\pm)}_{\bf p}(t) = e^{\mp i vp t/\hbar}
(1 \pm ({\bf p} \cdot \vec{\sigma})/p) \Psi_{\bf p} ,
\]
where $\vec{\sigma}=(\sigma_x,\sigma_y)$ is a vector of Pauli
matrices. A straightforward calculation of the current operator in
the Heisenberg picture \cite{Kat05} shows that an oscillatory
component in time exists due to an interference of
$\Psi^{(+)}_{\bf p}(t)$ and $\Psi^{(-)}_{\bf p}(t)$ solutions of
the DE. Therefore, in order to see a signature of ZB in the current,
it is important that electron-like and hole-like solutions interfere.
More generally, an oscillatory component of the current arises
if the electron field operator contains solutions to the DE at different
energies. Consequently, if we calculated the current operator
corresponding to a wave function solution to the DE with a fixed
energy $\ep$, then there would be no
oscillatory component of the current operator left and, thus, no
sign of ZB in the current. Importantly, this is the general situation in
ballistic transport in graphene if a plane wave solution (of a particle with
energy $\ep$) to the DE is injected from one region to another.
Thus, we conclude that ballistic dc transport in graphene shows no 
direct signature of ZB.

In principle, the Tien-Gordon setup of photon-assisted electron
transport for Dirac electrons, illustrated in Fig.~\ref{tg_setup}, 
can be used to generate the desired state.
The reason is that the
ac signal stimulates the absorption and emission of photons, which
in turn lead to a population of different side bands around the
Fermi energy. Then, the resulting electron field
operator in region $Gr(\rm in)$, Eq.~(\ref{psitr}), 
is precisely what is needed to
observe ZB in graphene. However, the detection of the resulting
oscillating current, Eq.~(\ref{curr1}), 
is not sufficient to prove the existence 
of ZB in graphene. The reason is that a preparation of a state that is a
superposition of different energy solutions to the free
Schr{\"o}dinger equation also yields a similar oscillating current 
(see, for instance, Eq.~(16) of Ref.~\cite{Ped98}). The only difference
to the free Schr{\"o}dinger case as compared to the Dirac case is the
peculiar energy dependence of the transmission in the latter.
Therefore, the Tien-Gordon setup analyzed in this paper is potentially
relevant to the detection of ZB, but our analysis indicates that the
truly fundamental signature of ZB needs to be identified more precisely.

\section{Conclusions}
\label{sec_con}

We have emphasized in this article that
the energy dependence of the transmission (typical for mesoscopic 
graphene systems) plays a crucial role for the unexpected transport
phenomena predicted in these devices, for recent examples see 
Refs.~\cite{Kat05,Two06,Che06,Kat06,Bee06,Tit06a,Cue06,Tit06b}. 
Therefore, it is desirable to directly probe
the energy dependence of the transmission. We have demonstrated that
photon-assisted transport is the natural way to do so. Furthermore, we
have pointed out that a possible application of our setup 
is to use graphene as a spectrometer for finite frequency
noise. Finally, we have discussed the relevance of photon-assisted
transport in graphene to the observability of ZB. Our conclusion is that 
photon-assisted transport can be used to create the conditions to observe ZB.
However, our analysis also shows that the truly fundamental signature of ZB 
(which is important for its detection) needs to be pinpointed in future 
studies.

We thank C.W.J. Beenakker, H. Heersche, P. Jarillo-Herrero, J. Schliemann,
I. Snyman, and L.M.K. Vandersypen for interesting discussions.
This research was supported by the Dutch Science Foundation NWO/FOM.


\begin{thebibliography}{99}
\bibitem{Foot1} A similar effect to the one we discuss in this article is 
seen if the ac signal is applied to the bias rather than to the gate. Both
ways are feasible from an experimental point of view.
\end{thebibliography}
\end{document}